\documentclass[aps,prd,reprint,groupedaddress]{revtex4-1}

\usepackage{graphicx}	
\usepackage{amsmath}	
\usepackage{amssymb}	

\usepackage{color}

\begin{document}

\title{Quantum-enhanced interferometry for axion searches}

\author{Denis Martynov}
\affiliation{University of Birmingham, Birmingham B15 2TT, United Kingdom}

\author{Haixing Miao}
\affiliation{University of Birmingham, Birmingham B15 2TT, United Kingdom}

\date{\today}

\begin{abstract}
We propose an experiment to search for axions and axion-like-particles in the galactic halo using quantum-enhanced interferometry. This proposal is related to the previously reported ideas (Phys. Rev. D 98, 035021, Phys. Rev. Lett. 121, 161301, Phys. Rev. D 100, 023548) but searches for axions in the mass range from $10^{-16}$\,eV up to $10^{-8}$\,eV using two coupled optical cavities. We also show how to apply squeezed states of light to enhance the sensitivity of the experiment similar to the gravitational-wave detectors. The proposed experiment has a potential to be further scaled up to a multi-km long detector. We show that such an instrument has a potential to set constrains of the axion-photon coupling coefficient of $\sim 10^{-18}$\,GeV$^{-1}$ for axion masses of $10^{-16}$\,eV or detect the signal.
\end{abstract}


\maketitle

\section{Introduction}

The challenge of discovering dark matter particles comes from a variety of candidates and their interaction with the Standard Model. Weakly interacting massive particles were the most promising dark matter candidates over the last few decades. However, a set of ultra-sensitive detectors, such as XENON~\cite{Aprile_Xenon_2017}, LUX~\cite{Akerib_Lux_2013}, and PandaX~\cite{Zhang_PandaX_2018} have not observed dark matter particles up to date and will reach neutrino background in near future~\cite{Baudis_2014}. Furthermore, the Large Hadron Collider has placed stringent constraints on supersymmetry that provids the theoretical basis for weakly interacting massive particles~\cite{Canepa_2019}. Therefore, it is important to diversity dark matter searchers. In this paper, we consider axions~\cite{Peccei_1977} and axion-like-particles~\cite{Graham_2013, Ringwald_2012, Ringwald_2014, Farina_2017} (ALPs) that are also well-motivated dark matter candidates.

Axions and ALPs are generically expected in many models of physics beyond the Standard Model~\cite{Svrcek_2006}. These particles emerge as the Goldstone bosons of
global symmetries that are broken at some high energy scale~\cite{Peccei_1977, Weinberg_1978, Dine_1981}.
If dark matter consists of ALPs with mass $m_a$ then its field behaves classically and can be written as~\cite{Budker_Casper_2014}
\begin{equation}
\label{eq:alps_field}
    a(t) = a_0 \sin(\Omega_a t + \delta(t)),
\end{equation}
where the angular frequency $\Omega_a = 2\pi f_a = m_a$ in the natural units ($\hbar=c=1$), $a_0 = \sqrt{2 \rho_{\rm DM}} / m_a$ is the amplitude of the field, $\rho_{\rm DM} \approx 0.3$\,GeV/cm$^3$ is the local density of dark matter, $\delta (t)$ is the phase of the field. The phase remains constant for times $t \lesssim \tau_a$, where $\tau_a = Q_a /f_a$ is the coherence time of the field, $Q_a = v^{-2} \sim 10^6$ is the quality factor of the oscillating field, and $v$ is the galactic virial velocity of the ALP dark matter~\cite{Graham_2013}. Eq.~(\ref{eq:alps_field}) neglects spacial variations of the field since ALPs wavelength $\lambda_a = (f_a v)^{-1} > 100$\,km  is significantly larger than the length of the proposed experiment for $m_a < 10^{-8}$\,eV.

The Goldstone nature of ALPs manifests itself in their derivative interactions with the Standard Model~\cite{Budker_Casper_2014}. In this paper, we consider an interaction of ALPs with photons parameterized by the coefficient $g_{a \gamma}$. The observable quantity is the phase difference accumulated by the left- and right-handed circularly polarized light that propagate in the presence of the ALPs field for a time period $\tau$. This phase difference is given by the equation~\cite{DeRocco_2018}
\begin{equation}
    \Delta \phi(t, \tau) = g_{a\gamma}[a(t) - a(t-\tau)]
    \label{eq:vel}
\end{equation}
and can be measured by sensitive laser interferometers~\cite{DeRocco_2018, Obata_2018, Liu_2019}. For $\tau = 10$\,nsec and $g_{a\gamma} = 10^{-10}$\,GeV$^{-1}$, we get the amplitude of $\Delta \phi$ equal $\approx 3.2 \times 10^{-15}$\,rad. This phase shift is significantly smaller compared to the ones observed by the gravitational-wave detectors, such as such as LIGO~\cite{LSC_aLIGO_2015} and Virgo~\cite{Acernese_aVIRGO_2015}. A typical source modulates the laser phase by $\sim 10^{-12} - 10^{-11}$\,rad~\cite{LSC_GW150914, LSC_GW151226, LSC_GW170104} but only lasts for a fraction of a second for $\sim 30$ solar mass black holes and $\approx 30$\,sec for neutron stars. However, dark matter signal can be accumulated during much longer times scales that are only limited by the duration of the experiment.

Recently, new configurations to search for ALPs were proposed in the literature~\cite{Kahn_2016, Ouellet_2019, DeRocco_2018, Nagano_2019, Obata_2018, Liu_2019}. Authors in~\cite{Nagano_2019} propose to search for axions around the free-spectral-range of linear cavities while authors in~\cite{DeRocco_2018} consider quarter-wave plates inside these resonators to search for axions at lower frequencies (below $\approx 20$\,kHz).  Authors in~\cite{Obata_2018} propose the design without the intracavity wave-plates by utilizing a bow-tie cavity with two counter-propagating beams. This detector has a potential to search for axions with masses below $10^{-12}$\,eV due to a limited bandwidth of the optical resonator. Some schemes were proposed in the literature~\cite{Miao_Unstable_2015, Miao_kHz_2018} to enhance the gain-bandwidth product of optical cavities but these schemes were not experimentally demonstrated yet. Authors in~\cite{Liu_2019} found a different approach to increase the range of axion masses in their proposal up to $10^{-8}$\,eV. They utilize a folded optical cavity with non-degenerate eigen P- and S-polarisation modes. The frequency difference between these polarisations is tuned to a particular axion mass. By changing the frequency between P- and S-pol in the optical cavity, the authors proposed to search for ALPs with masses from $10^{-13}$\,eV up to $10^{-8}$\,eV for $\sim 10$\,m long interferometers.

In this paper, we further advance the studies in~\cite{DeRocco_2018, Obata_2018, Liu_2019} and (i) propose a new optical configuration to scan for ALPs with masses from $10^{-16}$\,eV up to $10^{-8}$\,eV in Sec~\ref{sec:layout}, (ii) show how to enhance the sensitivity of our detector with squeezed states of light~\cite{LSC_SQUEEZING_2011, LSC_SQUEEZING_2013}, and (iii) calculate the sensitivity of the proposed detector with lengths of 2.5\,m and 4\,km to ALPs in Sec~\ref{sec:sens}. The former length is a typical scale of a table-top interferometer, while the later one is a scale of the gravitational-wave detectors, such as LIGO and Virgo. Once the third generation facilities, such as Einstein Telescope~\cite{Punturo_ET_2010} and Cosmic Explorer~\cite{LSC_FUTURE_2017}, are built, the current facilities have a potential to search for dark matter using optical interferometers. We summarise our conclusions in Sec~\ref{sec:conclusions}.

\section{Optical layout}
\label{sec:layout}

The proposed interferometer measures a difference in phase velocities between left- and right-handed circularly polarized light which propagates in the presence of the ALPs field. This effect can be equivalently understood as a slow rotation of the polarisation angle of a linearly polarised light in the ALPs dark matter~\cite{Liu_2019, Ivanov_2019}. Our detector consists of two folded optical resonators: main and auxiliary cavities as shown in Fig.~\ref{fig:layout}. The main optical cavity resonates a strong pump field in the horizontal polarisation (P-polarisation) which is partially converted to the vertical polarisation (S-polarisation) by the ALP field.

The first challenge is to amplify both P- and S-polarised fields with frequencies $\omega_p$ and $\omega_s$ in the main optical cavity given that optical frequencies of these fields are separated by the ALPs frequency $\Omega_a$. Similar to~\cite{Liu_2019} this challenge is solved by the folded design of the main cavity. Its two resonating modes are non-degenerate since P- and S- polarisations acquire different phases upon non-normal reflection from the cavity mirrors. The second challenge is to dynamically tune $\omega_p - \omega_s$ to scan for ALPs masses. Authors in~\cite{Liu_2019} propose to change angles of incidence of the laser beam on the cavity mirrors. Indeed, this approach will change the frequency separation between P- and S-polarisation but can also dramatically reduce a quality factor of the optical cavity making it insensitive to the axion field. Instead of changing the angles of incidence, we propose a coupled cavity design similar to the gravitational-wave detectors. The auxiliary cavity will allow us to dynamically tune $\omega_s$ and scan over a broad range of ALPs masses from $10^{-16}$\,eV up to $10^{-8}$ eV by detuning the auxiliary cavity from its resonance.

In this section, we discuss how the ALPs field produce the signal in the S-polarisaton in the main cavity, discuss how the auxiliary cavity tunes the eigen mode $\omega_s$ of the main cavity, and present the optical readout scheme.

\subsection{Propagation of the fields in the main cavity}

We now consider how linearly polarized light propagates in the ALPs field between two points separated by a distance $L$. We adopt Jones calculus with the electric field vector given by $(E_p, E_s)^T$, where $E_p$ and $E_s$ are the horizontal and vertical components of the field. The Jones matrix $P$ for propagation of light in the ALPs field is given by the equation
\begin{equation}
\label{eq:linear_propagation}
\begin{aligned}
   P
    &=
    A^{-1}
    \begin{pmatrix}
        e^{i\Delta \phi/2} & 0 \\
        0 & e^{-i\Delta \phi/2}
    \end{pmatrix}
     A \\
    &\approx 
    \begin{pmatrix}
        1 & \Delta \phi/2 \\
        -\Delta \phi/2 & 1
    \end{pmatrix},
\end{aligned}
\end{equation}
where matrices $A$ and $A^{-1}$ convert electric fields from the linear to circular basis and back. In Eq. (\ref{eq:linear_propagation}) we assume that $\Delta \phi \ll 1$ according to the discussed below Eq.~(\ref{eq:vel}). Eq. (\ref{eq:linear_propagation}) implies a slow rotation of the polarisation angle of a linearly polarised light in the ALPs field.

\begin{figure}
 \includegraphics[width=\columnwidth]{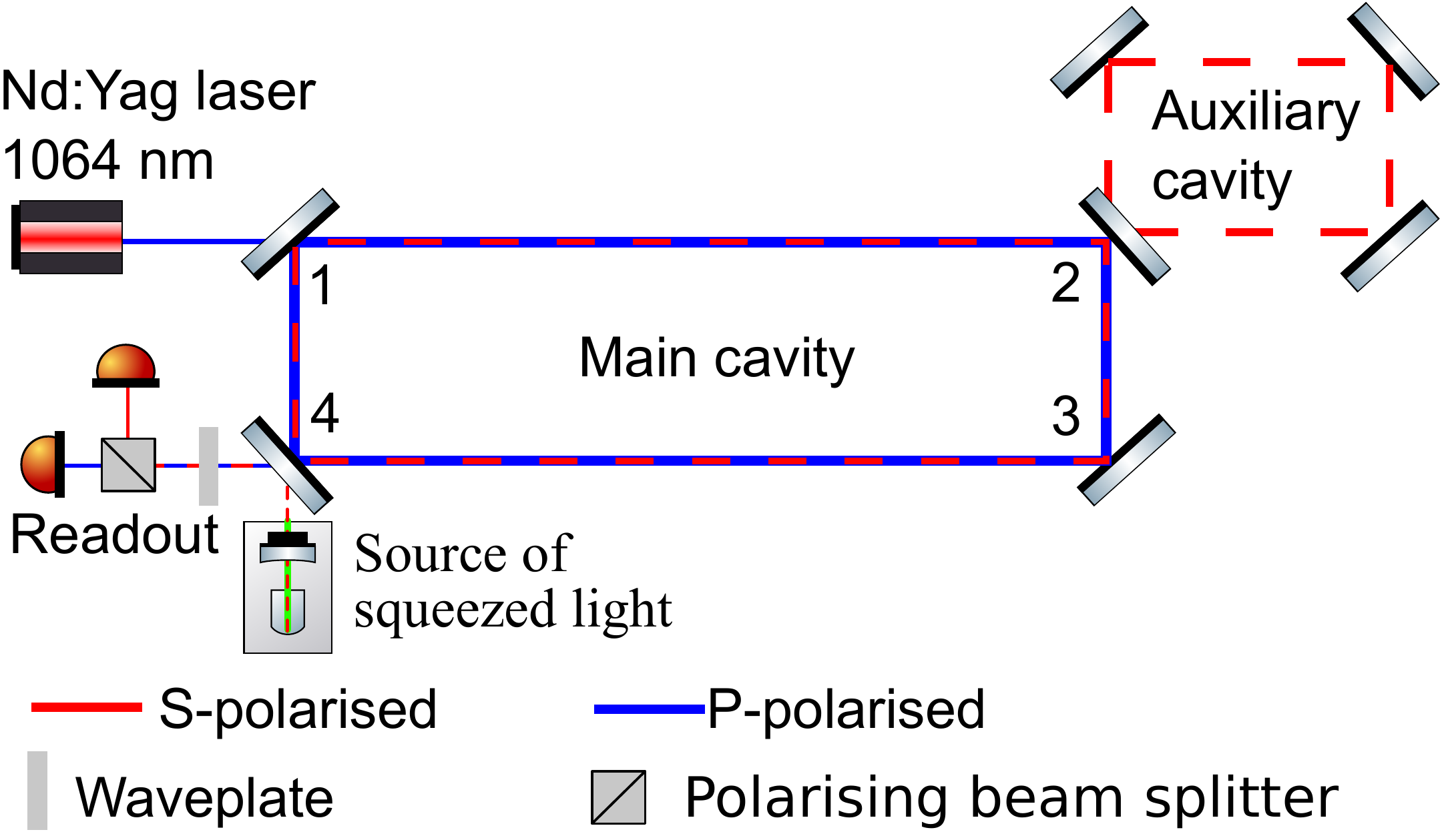}%
 \caption{Optical layout of the proposed experiment that consists of the main and auxiliary optical cavities. The mirrors in the main cavity have numbers 1 to 4 according to the discussion in the text.}
  \label{fig:layout}
\end{figure}

Phase shift $\Delta \phi$ is amplified in a high-finesse optical cavity. However, since P- and S- polarisation acquire phase difference of $\pi$ upon reflection from a mirror under normal angle of incidence, rotation of the polarisation angle will be cancelled after the round trip propagation of the field inside the linear optical cavity~\cite{DeRocco_2018}. Mathematically, the round trip Jones matrix for a linear cavity is $R_1 P(t) R_2 P(t-\tau/2) \approx I$ for $f_a \ll 1/\tau$, where $\tau$ is the cavity round trip time, $R_1=R_2=\bigl( \begin{smallmatrix}-1 & 0\\ 0 & 1\end{smallmatrix}\bigr)$ are Jones matrices for the mirrors at normal incidence and $I$ is a 2 by 2 identity matrix.

In order to accumulate $\Delta \phi$ over many bounces inside an optical cavity, we introduce folding in the cavity as shown in Fig.~\ref{fig:layout}. Distance between the mirrors 1 and 4 and mirrors 2 and 3 is significantly smaller compared to the distance between mirrors 1 and 2 and mirrors 3 and 4. Therefore, we can neglect any rotation of the polarisation angle between these mirrors by the ALPs field and the round trip Jones matrix is given by the equation
\begin{equation}
\label{eq:round_trip}
    Q = M_1 M_4 P(t) M_3 M_2 P(t-\tau/2),
\end{equation}
where matrices $M_1$, $M_2$, $M_3$, $M_4$ correspond to reflection of the laser field from each of the four mirrors. Matrix $M_2$ describes reflection of the laser light from auxiliary cavity. We can express Jones matrices of the mirrors as
\begin{equation}
    \begin{aligned}
        M_i = 
         \begin{pmatrix}
        -1 & 0 \\
        0 & e^{i \beta_i}
    \end{pmatrix},
    \end{aligned}
\end{equation}
where $\beta_i$ is the phase difference accumulated by the fields in P- and S-polarisation during the propagation inside the optical coatings.

In general, $\beta_i \neq 0$ since reflectively of each coating layer is different for P- and S-polarisation according to the Fresnel equations. This inequality leads to non-generate frequencies of the P- and S-polarised modes $\omega_p \neq \omega_s$. We propose to design stacks of the optical coating such that $e^{i (\beta_1 + \beta_4)} \approx 2 \pi K$ and $e^{i (\beta_2 + \beta_3)} \approx 2 \pi D$, where $K$ and $D$ are integer numbers. In this case, $M_1 M_4 = I$ and $M_2 M_3 = \bigl( \begin{smallmatrix}1 & 0\\ 0 & e^{i\beta}\end{smallmatrix}\bigr)$, where $\beta$ is an extra phase accumulated by the S-polarised beam inside the auxiliary cavity.



\subsection{Auxiliary cavity}

\begin{figure}
 \includegraphics[width=\columnwidth]{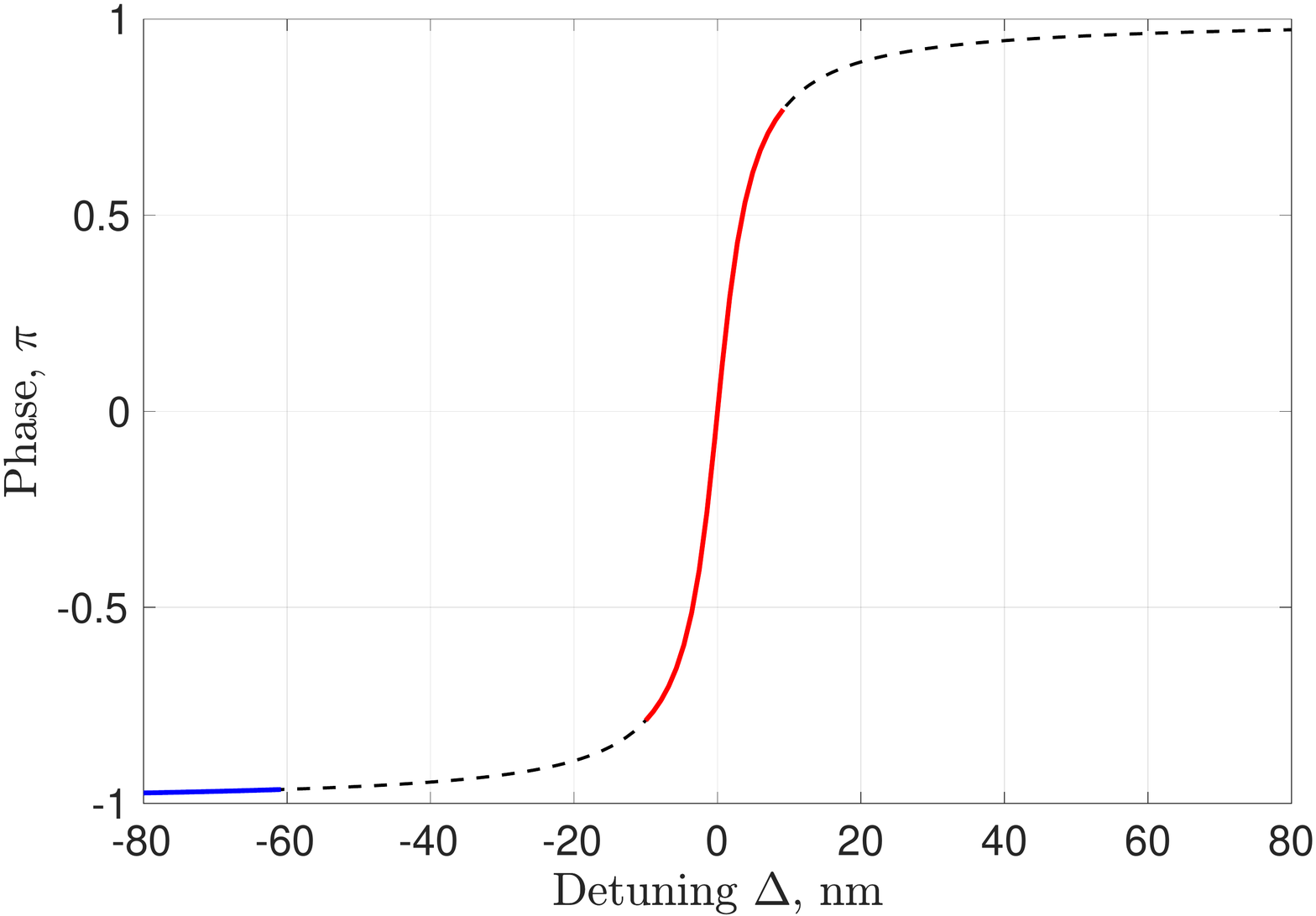}%
 \caption{Phase of the fields in P- and S-polarisation reflected from the auxiliary cavity. By detuning the auxiliary cavity by 20\,nm, we can tune the relative phase shift between the fields in P- (shown in blue) and S-polarisation (shown in red) by $\pi$.}
  \label{fig:reflectivity}
\end{figure}

We now discuss how the phase shift $\beta$ is dynamically tuned by the auxiliary cavity. If phases accumulated by the fields in P- and S-polarisation are $\xi_p$ and $\xi_s$ then the reflection coefficient from the auxiliary cavity is given by the equation~\cite{Black_2001}
\begin{equation}
    r_{p,s} = \frac{-r_2 + e^{i \xi{p,s}}}{1 - r_2 e^{i \xi_{p,s}}},
\end{equation}
where $r_2$ is the field reflectivity of the mirror 2. Fig.~\ref{fig:reflectivity} shows the argument of $r_{p,s}$ for different phases accumulated in the auxiliary cavity. We control this cavity such that $\xi_p \approx -\pi$. In this case, $r_p$ = -1 even for small changes of $\xi_p$. S-polarised beam is close to the resonance in the auxiliary cavity $\xi_s \ll 1$ and we can write the argument of $r_s$ as
\begin{equation}
\label{eq:beta}
    \beta = \frac{4}{T_2}\xi_s,
\end{equation}
where $T_2 = 1 - r_2^2$ is the power transmission of the mirror 2.
Small tuning of $\xi_s$ leads to significant phase shift of the field in the S-polarisation in the main cavity $\beta$. This phase can be controlled using an auxiliary laser which is phase-locked to the main laser~\cite{Staley_LOCK_2014}. We use this procedure to control the frequency of the eigen mode $\omega_s$ as discussed below.

\subsection{Optical readout}

We use Eq.~(\ref{eq:vel}) and~(\ref{eq:round_trip}) to calculate the field equations for the S-polarised field in the main cavity. In the further analysis we neglect the time dependence of the pump field in the cavity $E_{\rm p, cav}$ since it is not affected by the ALPs field. The field in the S-polarisation builds up in the main cavity due to the ALPs field according to the equation
\begin{equation}
\label{eq:resonating_fields}
        \begin{split}
        E_{\rm s, cav}(t) &= E_{\rm s, cav}(t-\tau)e^{i \beta}\sqrt{1-T_s} -\\
        &\frac{1}{2}[\Delta \phi(t, \frac{\tau}{2}) + \Delta \phi(t-\frac{\tau}{2}, \frac{\tau}{2})] E_{\rm p, cav},
        \end{split}
\end{equation}
where $T_s$ is the power transmission of the mirror 4 to the S-polarisation. The first term on the right-hand side of Eq.~(\ref{eq:resonating_fields}) represents the feedback term in the optical cavity while the second term is the excitation of the resonating field.

Eq.~(\ref{eq:resonating_fields}) can be solved in the frequency domain. We use a Fourier transform normalized to the coherence time of the ALPs field
\begin{equation}
    E(\Omega) = \frac{1}{\tau_a}\int_0^{\tau_a} E(t) \exp(-i\Omega t) dt
\end{equation}
and similar to~\cite{Budker_Casper_2014} we treat the ALPs phase $\delta$ from Eq.~(\ref{eq:alps_field}) as a constant over time period $\tau_a$. Solving Eq.~(\ref{eq:resonating_fields}) in the frequency domain, we get the solution in the form
\begin{equation}
\label{eq:s_field}
    \begin{split}
    E_{\rm s, cav}&(\Omega_a) = -\frac{E_{\rm p, cav}\exp(i\frac{\beta-\Omega_a \tau}{2}+\delta)}{1-\sqrt{1-T_s}\exp(i(\beta-\Omega_a \tau))}  g_{a\gamma} \times \\
    & \frac{\tau}{4} {\rm sinc} \left(\frac{\Omega_a \tau}{4} \right) \cos \left(\frac{2\beta-\Omega_a \tau}{4} \right)\sqrt{2 \tau_a \rho_{\rm DM}}.
    \end{split}
\end{equation}

\begin{figure}
 \includegraphics[width=\columnwidth]{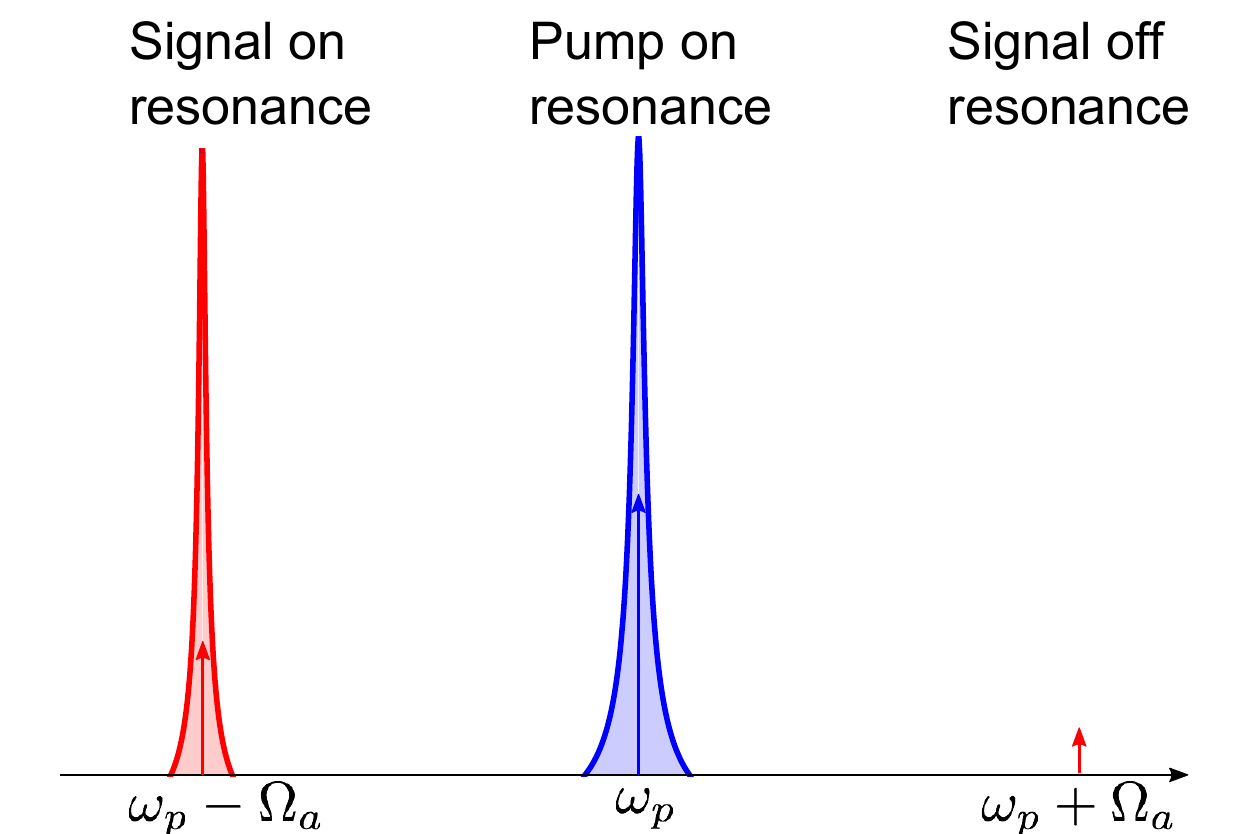}%
 \caption{ALP field converts the pump beam in the P-polarisation into the beam in S-polarisation. Both fields are resonantly enhanced in the main optical cavity.}
  \label{fig:p_s_resonance}
\end{figure}

The signal field in the S-polarisation is measured in transmission of the mirror 4 using the heterodyne readout. The field in the P-polarisation serves as a local oscillator in our readout scheme. First, we introduce a quarter wave plate to shift the phase between P-polarasation and S-polarisation by $\pi/2$ (see Fig.~\ref{fig:layout}). The latter shift will allow us to measure the phase quadrature of the field given by Eq.~(\ref{eq:s_field}). Then we introduce a half waveplate to convert a small fraction of the P-polarised light to S-polarised light $E_{\rm LO} = i\zeta E_{\rm p, cav} \sqrt{T_p}$, where $\zeta$ is twice the rotation angle of the half waveplate, $T_p$ is the power transmission of mirror 4 to the P-polarisation. Fourier transform of the power in the S-polarisation at the readout port is given by equation
\begin{equation}
\label{eq:signal_power}
        P_{\rm out}(\Omega_a) = E_{\rm LO} \sqrt{T_s}
        [E_{\rm s, cav}^*(-\Omega_a)-E_{\rm s, cav}(\Omega_a)],
\end{equation}
which implies that $P_{\rm out}(\Omega_a)$ is resonantly enhanced if the following condition is satisfied \begin{equation}
    \Omega_a = \pm \beta / \tau.
\end{equation}
Therefore, the ALPs mass equals the frequency separation of the P- and S-polarisation eigen modes in the main cavity that is determined by Eq.~(\ref{eq:beta}). Full-width at half-maximum of the resonance equals
\begin{equation}
\Delta f = \frac{T_s}{2\pi}{\rm FSR},
\end{equation}
where FSR is the free spectral range of the main cavity. Resonant amplification of the S-polarised light is schematically shown in Fig.~\ref{fig:p_s_resonance}.

\section{Sensitivity} 
\label{sec:sens}

Noise sources in the laser interferometers were actively studied in context of gravitational-wave detectors~\cite{LSC_aLIGO_2015, Acernese_aVIRGO_2015, Martynov_Noise_2016}, opto-mechanical setups~\cite{Cohadon_1999, Corbitt_2007, Purdy_2017, Sudhir_2017, Martynov_QUCORR_2017}, and laser gyroscopes~\cite{Hurst_2009, Schreiber_2011, Belfi:12, Korth_Gyro_15, Martynov_Gyro_2019}. Gravitational-wave detectors and opto-mechanical setups reach fundamental shot noise at frequencies above $\approx 40$\,Hz while at lower frequencies the sensitivity degrades due to ground vibrations, thermal noises, and scattered light~\cite{Tse_2019, Acernese_2019, Martynov_THESIS_2015, Gras_CTN_2017}. In the proposed experiment, the pump and signal fields follow the same path in the main cavity. Moreover, the mode of the auxiliary cavity has the same noise as the pump since the auxiliary cavity will be actively controlled with an auxiliary laser locking scheme~\cite{Staley_LOCK_2014}. Therefore, displacement noises in the main and auxiliary cavities will be cancelled out in the readout. The main source of the classical noises comes from intensity fluctuations of the pump beam. These fluctuations will be measured in transmission of the polarising beam splitter (see Fig.~\ref{fig:layout}) and fed back to the laser in a high bandwidth loop. In this paper, we calculate the sensitivity of the proposed experiment above 25\,mHz based on the quantum noise level.

\subsection{Quantum squeezing}

The main source of quantum noises comes from vacuum fluctuations which enter the interferometer from the readout port and through optical losses inside the cavity. Since we read out the phase quadrature of the field on the signal  photodetector, the optical power due to vacuum fields $b$ and $b^\dagger$ is given by the equation~\cite{Schumaker_TWOPHII_1985, Chen_QUANTUM_2013}
\begin{equation}
    P_{\rm shot}(\Omega_a) = E_{\rm LO} [b^\dagger(-\Omega_a) - b(\Omega_a)],
\end{equation}
where vacuum fields $b$ and $b^\dagger$ are in the S-polarisation and come from the open port of the interferometer, reflect from the mirror 4 and proceed to the signal photodetector. Therefore, we need to squeeze vacuum fields in the S-polarisation as shown in Fig.~\ref{fig:layout}. Then the power spectrum density of the shot noise is given by the equation
\begin{equation}
\label{eq:shot_power}
    |P_{\rm shot}(\Omega_a)|^2 = 2 \omega_p |E_{\rm LO}|^2 \exp(-2r),
\end{equation}
where $r$ is the squeezing factor. Modern optical table-top experiments reach $\exp(-r) \sim 0.15-0.5$~\cite{Vahlbruch_2016, Pradyumna_2018, Mehmet_2018} in the audio and radio-frequency bands and have the potential to reach a similar level of squeezing below 1\,Hz. In this paper, we use $\exp(-r) = 0.3$ for the proposed experiment.

The signal-to-noise (SNR) ratio of the setup for a particular ALPs mass $m_a$ is given by~\cite{Budker_Casper_2014}
\begin{equation}
\label{eq:snr}
    {\rm SNR^2} = \left| \frac{P_{\rm out}(\Omega_a)}{P_{\rm shot}(\Omega_a)} \right|^2 \sqrt{\frac{T_{\rm meas}}{\tau_a}},
\end{equation}
where $T_{\rm meas}$ is the measurement time. The latter multiplier comes from averaging the shot noise level around frequency $\Omega_a$ with a bandwidth of $\Omega_a / Q_a$ ~\cite{Budker_Casper_2014}. Eqs.~(\ref{eq:signal_power}) and~(\ref{eq:shot_power}) imply that $E_{\rm LO}$ cancels out in Eq.~(\ref{eq:snr}) and SNR does not depend on the level of the local oscillator field.

\subsection{Integration time}

\begin{figure}
 \includegraphics[width=\columnwidth]{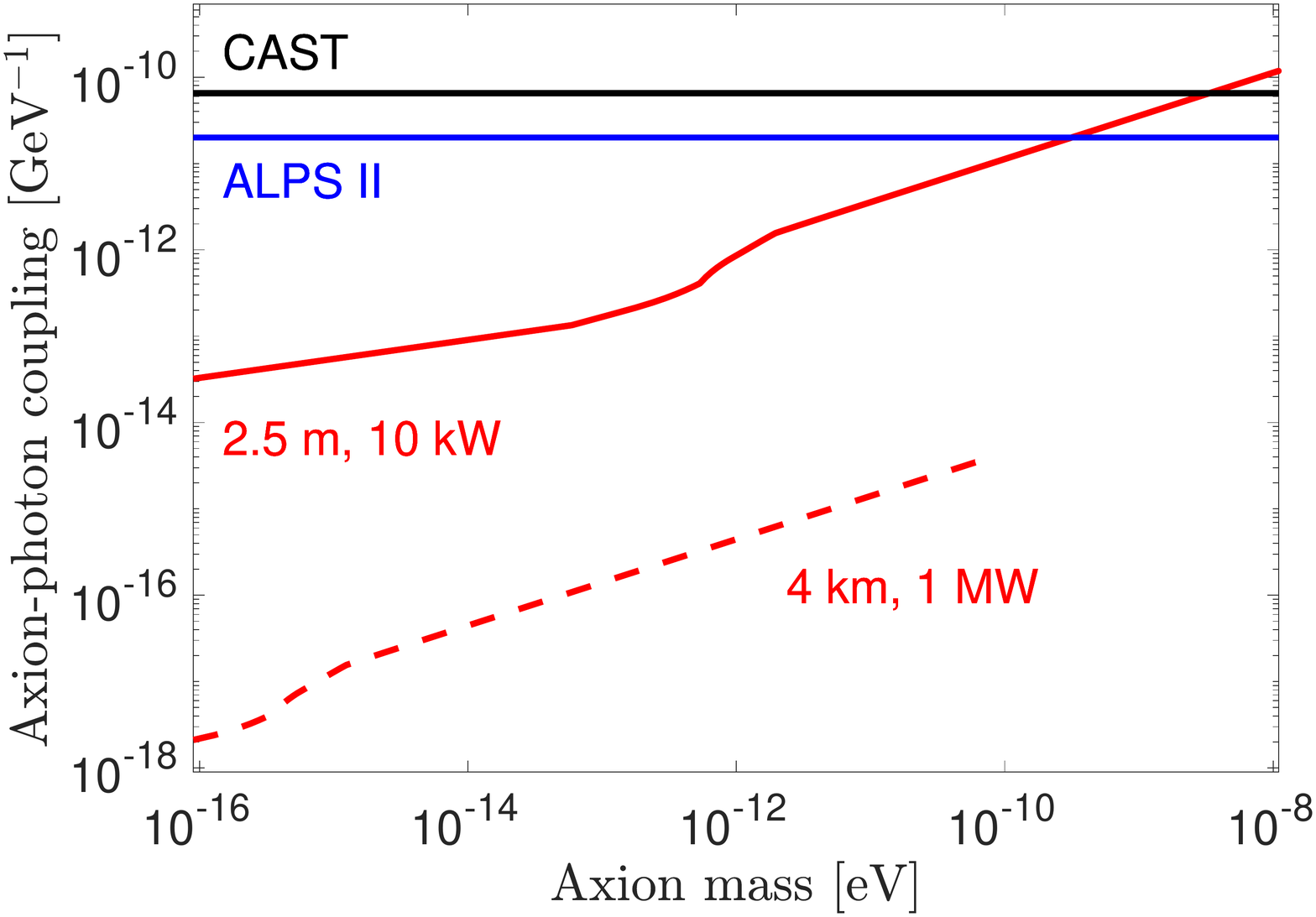}%
 \caption{Sensitivity of the proposed experiment to the axion-photon coupling coefficient after one year of integration and scanning through the ALP masses with signal-to-noise ratio of 1. We consider two optical configurations: a table-top setup with 10\,kW resonating power and a km-scale detector with 1\,MW of power. The distance shown in the figure refers to the distance between the mirrors 1 and 4. Existing limits from CAST~\cite{Anastassopoulos_2017} and design sensitivity of the ALPS II~\cite{Bahre_ALPS_2013} detector are shown for comparison.}
  \label{fig:sensitivity}
\end{figure}

We choose integration time for each ALP mass $T_{\rm meas} = N \tau_a$ according to~\cite{Liu_2019, Chaudhuri_2018, Khan_2016}. We scan over a range of ALPs masses by changing longitudinal offset of the auxiliary cavity in the range $0 \leq \beta \leq \pi/2$. Every step we shift the frequency difference between eigen modes of the main cavity by its full width half-maximum to keep resonance enhancement of the signal field in the S-polarisation. Given the total integration time of $T_{\rm int} = 1$\,year, the measurement time for a particular ALP mass in units of its coherence time is given by the equation
\begin{equation}
    N \approx \frac{\Delta f T_{\rm int}}{Q \ln ({\rm FSR}/\Delta f)},
\end{equation}
which implies that the SNR depends on the cavity finesse for the S-polarisation $\mathcal{F}_s = 2\pi / T_s$ according to the equation
\begin{equation}
\label{eq:snr_finesse_s}
    {\rm SNR} \sim g_{a\gamma} N^{1/4} \sqrt{\mathcal{F}_s} \sim g_{a\gamma} \left( \frac{\mathcal{F}_s}{\ln{\mathcal{F}_s}} \right)^{1/4}
\end{equation}
for $f_a > \Delta f$. In this paper, we consider $\mathcal{F}_s = 10^5$ and the latter multiplier in Eq.(\ref{eq:snr_finesse_s}) equals $\approx 10$. For $f_a < \Delta f$ the scaling of SNR with $\mathcal{F}_s$ is different from Eq.(\ref{eq:snr_finesse_s}) since the measurement time is determined by the coherence time of ALPs with masses $\pi \Delta f$. The SNR scales according to the equation
\begin{equation}
\label{eq:snr_finesse_s_band}
    {\rm SNR} \sim g_{a\gamma} \left( \frac{\mathcal{F}_s^2}{\ln{\mathcal{F}_s}}\right)^{1/4}
\end{equation}
for $f_a < \Delta f$ and the latter multiplier in Eq.(\ref{eq:snr_finesse_s_band}) equals $\approx 170$ for $\mathcal{F}_s = 10^5$.

The sensitivity of the experiment to the axion-photon coupling $g_{a\gamma}$ is shown in Fig.~\ref{fig:sensitivity} for SNR=1 and different lengths of the interferometer. The key property of this proposal is that the setup does not require strong magnets. Instead, ALPs dark matter converts the strong optical field in one polarisation into a field with an orthogonal polarisation. This property implies that the current gravitational-wave facilities can host dark matter detectors in the future. Both for the table-top 2.5\,m and long-scale 4\,km detectors the sensitivity curve significantly improves for $f_a < \Delta f$ since both modes $E_{\rm s, cav}(\Omega_a)$ and $E_{\rm s, cav}^*(-\Omega_a)$ resonate in the main optical cavity. These two fields interfere constructively on the signal photodetector since we measure the phase quadrature of the field. This interference explains the transition step in Fig.~\ref{fig:sensitivity} for the table-top experiment at $m_a \sim 10^{-12}$\,eV and for the km-scale experiment at $m_a \sim 10^{-15}$\,eV. Furthermore, the sensitivity of the experiment scales as $\tau_a^{-1/4} \sim m_a^{1/4}$ below $\Delta f$ according to Eqs.~(\ref{eq:s_field}) and~(\ref{eq:snr}) since the measurement time $T_{\rm meas}$ is the same for all frequencies smaller than $\Delta f$. This is contrast to larger masses ($\Omega_a > 2\pi \Delta f$) since we increase the measurement time as $T_{\rm meas} = N\tau_a$ and, therefore, the sensitivity scales as $\tau_a^{-1/2} \sim m_a^{1/2}$ above $\Delta f$ similar to the relation found in~\cite{Liu_2019}.

\section{Conclusions}
\label{sec:conclusions}

We proposed a new quantum-enhanced interferometer to search for ALPs in the mass range $10^{-16}$\,eV up to $10^{-8}$\,eV. This mass range corresponds to frequencies of the ALPs field from 25\,mHz up to 2.5\,MHz. In principle, the detector is sensitive at frequencies below 25\,mHz but we expect that the sensitivity will be limited by technical noises rather than by quantum noises similar to the optical gyroscopes~\cite{Hurst_2009, Schreiber_2011, Belfi:12, Korth_Gyro_15, Martynov_Gyro_2019}. An experiment is needed to measure the sensitivity at these low frequencies. The upper ALPs mass is limited by the free spectral range of the optical cavity and the corresponding sinc function in Eq.~(\ref{eq:s_field}). We proposed to scan over ALPs masses using an auxiliary cavity that tunes the frequency separation between the pump and signal fields in the main cavity.

We proposed a technique to enhance the quantum limited sensitivity of the interferometer by injecting squeezed states of light similar to the gravitational-wave detectors. Further steps include building a table-top prototype which can already improve over CAST limits in the ALPs mass range from $10^{-16}$\,eV up to $10^{-9}$\,eV. Once the technology is tested, the detector length can be scaled up. In particular, current gravitational-wave facilities are of significant interest to this proposal.

ALPs searches in the km-scale facilities have a potential to improve over the CAST limits by 5-7 orders of magnitude in the mass range $10^{-16}$\,eV up to $10^{-10}$\,eV or detect the signal. Current gravitational-wave facilities are ideal sites for axion interferometry since they already have vacuum envelopes and powerful lasers. New third generation gravitational-wave facilities will not only reach cosmological distances~\cite{Punturo_ET_2010, LSC_FUTURE_2017} but also create opportunities to use existing facilities for new experiments with dark matter. 

\begin{acknowledgments}
We thank Matthew Evans, Hartmut Grote, and Yuta Michimura for discussions of the optical configuration. We thank Henning Vahlbruch and Moritz Mehmet for discussions of sources of squeezed light. We also thank members of the 'QI' consortium Katherine Dooley, Stuart Reid, Animesh Datta, Robert Hadfield, Dmitry Morozov, and Vincent Boyer for useful discussions of the proposed experiment. D.M. and H.M. acknowledge the support of the Institute for Gravitational Wave Astronomy at University of Birmingham. H.M. is supported by UK STFC Ernest Rutherford Fellowship (Grant No. ST/M005844/11).
\end{acknowledgments}

\bibliography{Bibliography}

\end{document}